\documentstyle[aas2pp4]{article} 

\def\comment#1{{}}

\begin{document}

\title{Discovery of Microsecond Soft Lags in the X-Ray
Emission of the Atoll Source 4U~1636-536}

\author{Philip Kaaret\altaffilmark{1}, Santina
Piraino\altaffilmark{1,2,3}, Eric C. Ford\altaffilmark{4}, and
Andrea Santangelo\altaffilmark{2}}

\authoremail{pkaaret@cfa.harvard.edu}

\altaffiltext{1}{Harvard-Smithsonian Center for Astrophysics,
60 Garden St., Cambridge, MA 02138} 

\altaffiltext{2}{Istituto di Fisica Cosmica ed Applicazioni
dell'Informatica, CNR, Via Ugo La Malfa 153, 90146 Palermo,
Italy}

\altaffiltext{3}{Istituto di Fisica, Universita' di Palermo,
via Archirafi 36, 90123 Palermo, Italy } 

\altaffiltext{4}{Astronomical Institute, ``Anton Pannekoek'',
University of Amsterdam, Kruislaan 403, 1098 SJ Amsterdam,
The Netherlands}

\begin{abstract}

Exploiting the presence of kilohertz quasi-periodic
oscillations (QPOs) in the timing power spectrum, we find that
the soft x-ray emission of the neutron-star X-ray binary and
atoll source 4U~1636-536 modulated at the QPO frequency lags
behind that of the hard x-ray emission.  Emission in the
3.8--6.4~keV band is delayed by $25.0 \pm 3.3 \, \rm \mu s$
relative to the 9.3--69~keV band. The delay increases in
magnitude with increasing energy.  Our results are consistent
with those of Vaughan et al.\ (1997), when the sign is
corrected (Vaughan et al.\ 1998), for the atoll source
4U~1608-52.  The soft lag could be produced by Comptonization
of hard photons injected into a cooler electron cloud or by
intrinsic spectral softening of the emission during each
oscillation cycle.

\end{abstract}


\section{Introduction}

Millisecond quasi-periodic oscillations (QPOs) in the x-ray
emission of x-ray binaries are likely related to physical
processes occurring near the compact object and may, thus,
provide information about accretion flow in the strong
gravitational fields near the compact object (for a brief
review, see Kaaret \& Ford 1997).   The existence of the QPOs
can be exploited to measure additional properties of the x-ray
emission mechanisms and geometry.  Here, we use kHz QPOs with
relatively high coherence, $\nu / \Delta \nu > 100$, to
measure microsecond lags between emission in different x-ray
energy bands.

In \S 2, we describe the observations and technique used to
calculate the time lags.  In \S 3, we present the results of
the lag measurements. We conclude, in \S 4, with brief
comments on the physical implications.

\section{Observations and Analysis}

The following results use data from the Proportional Counter
Array (Zhang et al.\ 1993) on the Rossi X-ray Timing Explorer
(Bradt et al.\ 1993) with 122~$\mu$s time resolution for
x-rays in the 2--70 keV band.  The data presented here for
4U~1636-536 are from a 23~ks interval beginning UTC 1998 Feb
24 23:26:20 during which QPOs were detected.

Before attempting to measure time delays, we investigated the
QPOs.  We performed Fourier transforms on 1~s segments of data
with no energy selection, yielding a Nyquist cutoff frequency
of 4096~Hz and a transform window function with a width of
approximately 1~Hz.  The power spectra within 128~s intervals
were summed.  The total power spectrum for each 128~s interval
was searched for QPO peaks above 100~Hz.  A function,
consisting of a Lorentzian plus a constant, was fitted for
each trial frequency, defined as those points with power more
than $2 \sigma$ above the average.  All 128~s intervals for
which QPO peaks were detected with greater than $2 \sigma$
significance were retained.  In the case of multiple QPO
detections in one interval, the QPO with highest significance
was chosen. Typically, the QPOs were detected at better than
$4 \sigma$ significance.

\begin{figure}[tb] \epsscale{0.7} \plotone{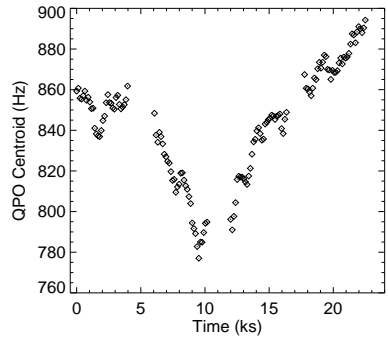}
\caption{Centroid of the QPO peak versus time for
4U~1636-536.  The time zero is UTC 1998 Feb 24 23:26:20.}
\label{fig_centroid} \end{figure}

The single QPO peak had a centroid varying in the range of
775--895~Hz, see figure~\ref{fig_centroid}.  The centroid
frequency could be determined to an accuracy better than
0.4~Hz.  The QPO peak widths had an average of 4.7~Hz.  The
intrinsic variation in the centroid is far larger than the
centroid measurement error or the QPO peak width.  

In using an oscillation (whether coherent or quasiperiodic) to
measure a time delay between different energy bands,  the
cross spectrum is integrated over a selected frequency range.
A narrow frequency range reduces the noise in the measurement,
but the range must be sufficiently broad to contain the
oscillation power.  In the standard approach, a fixed
frequency range is used for the full duration of an
observation (e.g. Vaughan et al.\ 1997).  Due to the large
variations in the QPO centroid in our observation, we chose
instead to integrate the cross spectrum over a range varying
with the QPO centroid.  We selected an integration range of
approximately twice the average QPO peak width.  Compared to
the fixed frequency range of 120~Hz required to contain the
wandering QPO peak, the signal to noise ratio for measurement
of a time delay in our observation of 4U~1636-536 is improved
by roughly a factor of 3.

We calculated the cross spectrum between a pair of energy
bands for each 1~s of data and found the average and variance
for each 128~s interval.  A complex cross correlation vector
was calculated for each 128~s interval by integrating over a
10~Hz frequency range centered on the QPO peak centroid found
for that interval.  We summed the complex cross correlation
vectors found for each interval to derive a complex cross
correlation vector for the whole observation.  The phase lag
given by the argument of this total complex cross correlation
vector was then converted to a time delay using the average of
the QPO centroid frequencies found in the 128~s intervals. 
The error was calculated from the variance in the real and
imaginary components of the  total cross correlation vector. 
Intrinsic as well as statistical fluctuations contribute to
the error estimate.  We also estimated the time delay for the
whole observation by averaging time delays calculated for
each 128~s interval.  The two methods gave consistent results,
which is expected since the fractional range of the QPO
centroid variation is relatively small, about 14\%.  We cannot
distinguish whether the lags remains constant in phase or in
time as the centroid frequency varies.  Below, we quote the
time delay found from the total cross correlation vector.

To determine the correct sign for the lags, we tested our
analysis code using simulated data and observations of
SAX~J1808.4--3658 for which the correct sign can be determined
by direct examination of the pulse-folded light curve in the
various energy bands (Cui, Morgan, \& Titarchuk 1998).

\section{Results}

\begin{figure}[tb] \epsscale{0.7} \plotone{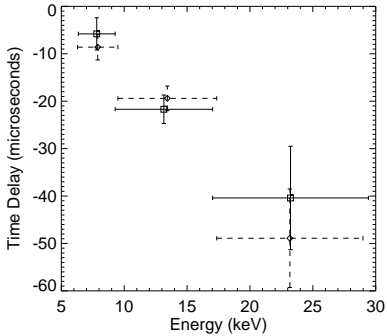}
\caption{Time delay versus energy for 4U~1636-536 (squares)
and 4U~1608-52 (diamonds).  The error bar on energy indicates
the band used to calculated the delay. The delays are relative
to the 3.8--6.4~keV band for 4U~1636-536 and to the
3.9--6.3~keV band for 4U~1608-52. Following Cui et al.\
(1998), the negative values emphasize the fact that the hard
X-rays actually lead the soft X-rays.} \label{fig_delay}
\end{figure}

Time delays, for 4U~1636-536, were calculated from 138
segments of 128~s each in which QPOs were detected.  The soft
emission in the 3.8--6.4~keV band lags the hard emission above
9.3~keV by $25.0 \pm 3.3 \, \rm \mu s$.  The coherence,
calculated following the prescription of Vaughan \& Nowak
(1997), is consistent with unity and is above 0.75. The
detection of the lag is significant at the $7 \sigma$ level.
The lag for the energy bands 6.4--9.3--17.0--29.4~keV versus
the 3.8--6.4~keV band are shown in figure~\ref{fig_delay}.  We
indicate the lags as negative to emphasize that the hard
emission precedes the soft emission.  The magnitude of the
delay increases with energy.

We also divided the data into the four contiguous segments of
QPO detections apparent in figure~\ref{fig_centroid}.  We find
no strong evidence for variability of the time delay between
any two segments.  The most significant difference is only
$1.6 \sigma$.

To check our analysis and also to have a second source for
comparison, we analyzed data for 4U~1608-52 for a 13~ks
interval beginning UTC 1996 Mar 3 19:18:20 extracted from the
public archive.  The 4U~1608-52 data show a single QPO varying
over the range 820--890~Hz.  We find the same sign and similar
magnitudes of delay in 4U~1608-52 as in 4U~1636-536.  We
detected QPOs in 66 segments of 128~s each and used an
integration width of 12~Hz to better match the QPO width in
4U~1608-52.  The soft emission in the 3.9--6.3~keV band lags
the hard emission above 9.5~keV by $21.4 \pm 2.7 \, \rm \mu
s$.  The energy dependence of the lag is shown in
figure~\ref{fig_delay}.  Delays are shown for the energy bands
6.3--9.5--17.4--29.0~keV relative to the 3.9--6.3~keV band. 
The energy bands were selected to best match those used in the
4U~1636-536 analysis, allowing for gain changes in the PCA. 
Vaughan et al.\ (1997) found delays of similar magnitude, but
opposite sign for part of this same observation.  The sign was
later corrected (Vaughan et al.\ 1998) and is now consistent
with our results.

\section{Discussion}

The facts that the hard emission precedes the soft emission
and that the modulation of the QPO is higher in the hard band
(Zhang et al. 1996) suggest that the process which generates
the QPO acts earlier and more strongly at higher energies. 
The soft lag could be produced either by reprocessing of the
hard emission into a softer band or by spectral softening of
the intrinsic emission during each oscillation cycle. 

The lag is opposite in sign to that generally expected in
Comptonization models, in which the spectrum is usually
generated by upscattering soft photons (Sunyaev \& Titarchuk
1980), but could be due to Compton scattering of hard photons
injected into a scattering medium which is cooler than the
injected photon spectrum.  This mechanism has been suggested
for the x-ray millisecond pulsar, SAX~J1808.4--3658 (Cui et
al.\ 1998).  Using the delay between the 3.8--6.4~keV and
9.3--17.0~keV bands leads to an estimate (Cui et al. 1998) of
$\sim 0.1 \tau \, \rm km$ for the size of the Comptonization
region with optical depth $\tau$.  For $\tau \sim 12$, as
typically found for 4U~1636-536 (Christian \& Swank 1996), the
Comptonization region would be very compact, at most a few
kilometers.

Additional observations are required to measure more
accurately the variation of the lags with energy in order to
determine if the soft lags are generated by Comptonization or
by intrinsic softening of the emission on each QPO cycle. 
Also more realistic theoretical modeling, including generation
of the input hard photons, is required to adequately interpret
the observations.

\acknowledgments

P.~K. and S.~P. acknowledge support from NASA grants NAG5-7405
and NAG5-7334.



\end{document}